\documentclass[%
 reprint,
superscriptaddress,
 amsmath,amssymb,
 aps,
]{revtex4-2}

\usepackage{graphicx}
\usepackage{dcolumn}
\usepackage{bm}
\usepackage[section]{placeins}

\usepackage{color,soul}

\newcommand{\ket}[1]{\left| {#1} \right>}
\newcommand{\bra}[1]{\left< {#1} \right|}

\newcommand{\bq}{\mathbf{q}}
\newcommand{\br}{\mathbf{r}}
\newcommand{\bep}{\boldsymbol{\epsilon}}

\newcommand{\da}{\downarrow}

\newcommand{\rEq}[1]{Eq.~(\ref{#1})}
\newcommand{\eSI}{I}
\newcommand{\angstrom}{\mbox{\normalfont\AA}}

\begin{document}

\preprint{APS/123-QED}

\title{Attosecond imaging of photo-induced dynamics in molecules using time-resolved photoelectron momentum microscopy}

\author{Marvin Reuner}
\affiliation{I. Institute for Theoretical Physics and Centre for Free-Electron Laser Science, Universit{\"a}t Hamburg, Notkestraße 9, 22607 Hamburg, Germany}

\author{Daria Popova-Gorelova}
\affiliation{I. Institute for Theoretical Physics and Centre for Free-Electron Laser Science, Universit{\"a}t Hamburg, Notkestraße 9, 22607 Hamburg, Germany}
\affiliation{The Hamburg Centre for Ultrafast Imaging (CUI), Luruper Chaussee 149, 22607 Hamburg, Germany}

\date{\today}

\begin{abstract}
    We explore the novel capabilities offered by attosecond extreme ultraviolet and x-ray pulses that can be now generated by free-electron lasers and high-harmonics generation sources for probing photon-induced electron dynamics in molecules. 
    We theoretically analyze how spatial and temporal dependence of charge migration in a pentacene molecule can be followed by means of time-resolved photoelectron microscopy on the attosecond time scale. 
    Performing the analysis, we accurately take into account that an attosecond probe pulse leads to considerable spectral broadening. 
    We demonstrate that the excited-state dynamics of a neutral pentacene molecule in the real space map onto unique features of photoelectron momentum maps. 
\end{abstract}

\maketitle

\section{Introduction}
Photoelectron momentum microscopy is the technique, in which a light pulse ionizes a probe leading to the detachment of a photoelectron. The momentum distribution of the detached photoelectron encodes information about the shape of an orbital or a bond in the real space from which the electron was detached \cite{doi:10.1126/science.1169183, puschnig_reconstruction_2009, doi:10.1073/pnas.1315716110, PhysRevA.89.021403, PuschnigJESRP15, weis_exploring_2015, Popova-GorelovaPRA16, Kliuiev_2016, doi:10.1126/science.aar4183, Kliuiev2019, metzger_plane-wave_2020, Jansen_2020}. This connection is facilitated by the Fourier transform, which links the real and momentum space. The spatial resolution of this technique is given by the de-Broglie wavelength of the emitted photoelectron, which is determined by the kinetic energy of the photoelectron. If an extreme ultraviolet (XUV) probe pulse detaches an electron from a valence orbital, the photoelectron would also have a kinetic energy in the XUV range corresponding to the resolution of a few Angstroms.

In recent years, there has been a remarkable progress in atomic-scale imaging of photo-excited states and dynamics using photoelectron momentum microscopy. Momentum-space distribution of transiently excited electrons was measured \cite{doi:10.1126/science.abf3286}, the dynamics of crystalline bonds have been followed during the photo-induced phase transition \cite{doi:10.1126/science.aar4183} and the technique allowed to detect singlet fission with orbital resolution \cite{https://doi.org/10.48550/arxiv.2204.06824}. 
Time-resolved photoelectron momentum microscopy probing photo-induced dynamics of pentacene films on sub-picosecond time scale has been achieved at the free-electron laser FLASH \cite{baumgartner2022ultrafast}. So far, these experiments were performed at a sub-picosecond time scale and were not capable of capturing pure electron dynamics in real time on its natural time scale. This barrier is now feasible to overcome due to the new capabilities to produce attosecond x-ray pulses at free-electron lasers \cite{Hartmann2018, Maroju2020, duris2020tunable} or attosecond XUV pulses using high-harmonics generation sources \cite{Teichmann2016, Li2017, Ren_2018, Li2020, Rossi2020, doi:10.1063/5.0020649, Midorikawa2022}. 
These sources have already been successfully employed to study electronic processes with few-femtosecond to attosecond time and atomic-scale space resolution \cite{doi:10.1126/science.1254061, doi:10.1126/science.aab2160, von2018conical, Langer2018, Siegrist2019, Peng2019, Marangos2020, doi:10.1126/science.abf1656, PhysRevX.11.031001, Garratt2022, M_nsson_2022}, and new experimental schemes are being explored theoretically \cite{doi:10.1073/pnas.1202226109, PhysRevB.91.184303, PhysRevB.92.184304, PhysRevA.95.053411, PhysRevLett.121.203002, PhysRevLett.120.133204, doi:10.1021/acs.jpca.8b12313, doi:10.1063/1.5109867, doi:10.1063/4.0000016, doi:10.1073/pnas.2105046118, PhysRevResearch.4.013073, doi:10.1021/acs.jctc.2c00064, https://doi.org/10.48550/arxiv.2203.02698}. 
The interpretation of a signal from attosecond probe pulses is challenging, since they have a broad bandwidth in the energy domain, which smears out spectral lines in a signal. In this article, we analyze how attosecond XUV pulses can be employed to measure charge migration in a pentacene molecule and analyze how to extract the real-space and real-time information about charge migration using the momentum-resolved photoelectron microscopy.

Pentacene is a prototypical organic semiconductor composed of pentacene molecules, which is among candidate materials for efficient organic photovoltaics \cite{doi:10.1063/1.1829777, leo2016organic}. The understanding of the process of solar-energy conversion in organic semiconductors strongly relies on the insight into both intra- and intermolecular photo-excited processes \cite{kohler2015electronic}. Exciton migration is one of the fundamental processes governing solar-energy conversion and the role of electronic coherence processes becomes especially relevant near conical intersections \cite{doi:10.1098/rsif.2013.0901, Nelson2018, Rafiq2019}. Electronic charge-migration dynamics launched as a coherent superposition of electronic states have already been studied in detail \cite{CEDERBAUM1999205, doi:10.1063/1.1540618, 10.1007/978-3-540-95946-5_190, doi:10.1063/1.2970088, doi:10.1063/1.3353161, doi:10.1063/1.3506617, doi:10.1021/jz200887k, PhysRevA.83.013411, Kuleff_2014, doi:10.1063/1.4996505}. In this paper, we focus on the description of how such coherent electron dynamics can be probed on nanoscale by broadband probe pulses. Particularly, we study how attosecond photoelectron momentum microscopy can be employed to probe intramolecular coherent exciton dynamics in pentacene. 

It has been shown that time-resolved photoelectron spectroscopy can be applied to probe ultrafast electronic and nuclear molecular dynamics \cite{doi:10.1021/jp407295t, doi:10.1021/cr020683w, doi:10.1021/acs.jpclett.7b00877, doi:10.1021/acs.jpca.8b12313, Marciniak2019, doi:10.1021/acs.jpclett.1c01843, D2CP02725A}. However, angle-unresolved photoelectron spectra do not carry any spatial information about the dynamics, which is lost due to angle averaging. In photoelectron momentum microscopy, momentum distribution of photoelectrons is detected. The great advantage of this technique over angle-averaged spectroscopy is that the distribution provides details of electron dynamics within a molecule, if photoelectrons are detached with high kinetic energies. We will also show that photoelectron spectra can loose important time-resolved information about the dynamics after angle averaging.

In this article, we apply the general theoretical formalism to describe the time and momentum-depended photoelectron probability by a broad-bandwidth probe pulse derived in Ref.~\cite{Popova-GorelovaPRA16}. That study proposed to probe hole dynamics in a positively ionized molecule by emitting electrons from orbitals that were initially occupied. Here, we propose to apply attosecond XUV pulses to probe photo-excited states of a neutral molecule, which is especially relevant to reveal information about exciton dynamics in the energy conversion processes.

The article is organized as follows. We develop the theoretical description of time-resolved photoelectron momentum microscopy of excited-state dynamics in molecules taking electron correlations and broad-bandwidth of a probe pulse into account in Sec.~\ref{sec:method}. 
In Sec.~\ref{sec:dynmaicsPentacene}, we calculate photoelectron spectra and photoelectron momentum maps from coherent electron-hole dynamics in a pentacene molecule obtained by an attosecond XUV probe pulse. We reveal how time-resolved atomic-scale changes in electron density are correlated with time-resolved changes in photoelectron momentum maps.

\section{Method} \label{sec:method}
Throughout this article, we consider a molecule with coherently evolving electron dynamics. In this case, the $N_{\text{el}}$-electron wave function of a molecule is a coherent superposition of several excited electronic eigenstates $\ket{\Phi_\eSI}$
\begin{equation} \label{eq:cohState}
    \ket{\Psi(t)} = \sum_{\eSI} C_{\eSI} e^{-i E_{\eSI} (t - t_0)} \ket{\Phi_\eSI},
\end{equation}
where $E_{\eSI}$ is the corresponding eigenenergy and the coefficient $C_{\eSI}$ determines the population distribution for each eigenstate. The coherent superposition has emerged at time $t_0$ due to crossing a conical intersection or the excitation by a pump pulse. We assume that an ultrashort XUV probe pulse interacts with the molecule at time $t_p$, which leads to an emission of a photoelectron. The momentum distribution of the photoelectron encodes information about the electronic state at the time $t_p$. In our study, we propose to probe electron dynamics by studying the momentum distribution of photoelectrons emitted from outermost orbitals that became populated due to the photoexcitation of the molecule. We use atomic units for this and following expressions.

Assuming a probe pulse with the central photon energy $\omega_{\text{in}}$, polarization $\bep_{\text{in}}$, pulse duration $\tau_p$ and intensity profile $I(t) = I_0 e^{-4 \ln{2} [(t-t_p)/\tau_p]^2}$, the general expression for the photoelectron probability is \cite{Popova-GorelovaPRA16}

\begin{align} \label{eq:photoElProp}
\begin{split}
    P(\bq,t_p)= &
    \frac{\tau_p^2 I_0 \left|\bep_{\text{in}} \cdot \bq \right|^2}{8 \pi \ln2 \omega_{\text{in}}^2 c} 
    \sum_{F,\sigma} e^{-(\Omega_F - \epsilon_e )^2\tau_p^2/(4\ln2)} \\
    & \times \left| \chi_\sigma^\dagger
    \int\mathrm{d}^3 r\
    e^{-i\bq\cdot \br} 
    \phi_F^D(\br, t_p)
    \right|^2,
    \end{split}
\end{align}
where $\bq$ is the momentum of the photoelectron with the corresponding energy $\epsilon_e = \frac{|\bq|^2}{2}$ and $\chi_\sigma$ is the photoelectron spin state. The summation is over all possible final states of the molecule $\bra{\Phi_F^{N_{\text{el}}-1}}$ that can be produced after the photoionization by the probe pulse, $E_F$ is their corresponding eigenenergy. $\phi_F^D(\br,t_p) = \bra{\Phi_F^{N_{\text{el}}-1}} \hat{\psi}(\br) \ket{\Psi(t_p)}$ is the Dyson orbital defined as the overlap between the $N_{el}$ wavefunction of the initial state of the system and the $(N_{el}-1)$ wavefunction of the final state, where $\hat{\psi}(\br)$ is the electron annihilation field operator. 
$\Omega_F = \omega_{\text{in}} + \left<E\right> - E_F$, where $\left<E\right>$ is the mean energy of the eigenstates involved in the wave-packet in \rEq{eq:cohState}. Here, we assumed the sudden approximation, {\it i.~e.} the molecular state after photoionization and the state of the photoelectron are decoupled, which is justified for high kinetic photoelectron energies \cite{hufner2007very}. It is also assumed that the probe-pulse duration is much shorter than the characteristic time scale of the electron dynamics. This means that the energy splittings of the eigenstates involved in the dynamics are much smaller than the bandwidth of the probe pulse and can be substituted by their mean energy. If this assumption fails, the eigenenergies must enter the expression explicitly and Eq.~(\ref{eq:photoElProp}) gets slightly modified
\begin{align}
P(\bq,t_p) \propto\sum_{F,\sigma}\Biggl|&\sum_I e^{-(\omega_{\text{in}} + E_I - E_F - \epsilon_e )^2\tau_p^2/(8\ln2) }\label{Eq_long_pulse_duration}\\
&\times\chi_\sigma^\dagger
    \int\mathrm{d}^3 r\
    e^{-i\bq\cdot \br} 
    \phi_{FI}^D(\br, t_p)\Biggr|^2,\nonumber
\end{align}
where $\phi_{FI}^D(\br,t_p) = \bra{\Phi_F^{N_{\text{el}}-1}} \hat{\psi}(\br) \ket{C_{\eSI} e^{-i E_{\eSI} (t _p- t_0)} }$. Time- and angle-resolved photoelectron probability for molecules with electron dynamics has also been previously theoretically studied \cite{PhysRevA.86.053429, doi:10.1021/jp407295t, doi:10.1021/jp508218n, Mignolet_2014}. {Eqs.~(\ref{eq:photoElProp}) and (\ref{Eq_long_pulse_duration}) take a broad bandwidth of a probe pulse accurately into account, which was not the case in these studies.

It follows from the expression \rEq{eq:photoElProp} that the momentum-resolved photoelectron probability encodes the Fourier transform of the Dyson orbital. If the molecule is in the ground state, the overlap integral between a singly ionized molecule and the molecule before the ionization is well approximated by the Hartree-Fock (HF) orbital, from which an electron was detached. Due to this connection, photoelectron momentum microscopy is used as an orbital imaging method \cite{doi:10.1126/science.1176105, PhysRevB.84.235427, Stadtm_ller_2012}.

When a molecule in an excited state is ionized by a probe pulse, the Dyson orbital cannot be approximated by a molecular orbital and the interpretation of the Dyson orbital becomes more challenging. In order to carefully describe the excited state dynamics as well as accurately treat the Dyson orbital, we describe the excited states involved in the dynamics $\Phi_\eSI$ and the states of the ionized molecule $\Phi_F^{N_{\text{el}}-1}$ using the configuration interaction method. We express each excited state based on the configuration interaction approach limited to single excitations as

\begin{equation} \label{eq:cisExcitedState}
    \ket{\Phi_\eSI} = \sum_{i_1,a_1} \tilde{c}_{i_1}^{a_1}(\eSI) \ket{\phi_{i_1}^{a_1,S(I)}},
\end{equation}

where $\ket{\phi_{i_1}^{a_1,S(I)}}$ denotes a configuration state function (CSF), where an electron of the HF ground state has been excited from the orbital $i_1$ to the orbital $a_1$. 
Such a state has a hole in orbital $i_1$ and an electron in orbital $a_1$, and $S(I)$ describes the total spin and the spin projection value. For the description of the final state, we use a similar approach, but add one additional hole in the configuration space. 
Thus, the final state is
\begin{equation}
\begin{split}
    \bra{\Phi_F^{N_{\text{el}}-1}} = \sum_{i_2,j_2,a_2} & \tilde{c}_{i_2,j_2}^{a_2}(F) \bra{\phi_{i_2,j_2}^{a_2,S(F)}}\\
    & + \sum_{i_2} \tilde{c}_{i_2}(F) \bra{\phi_{i_2}^{S(F)}},
\end{split}
\end{equation}
where $\bra{\phi_{i_2,j_2}^{a_2,S(F)}}$ describes a CSF which contains two holes in orbitals $i_2$ and $j_2$ and an electron in orbital $a_2$. 
We denote an $N_{\text{el}}-1$-electron CSF with one hole among originally occupied orbitals of the molecule (HOMO, HOMO-1, ...) and no electrons in the originally unoccupied molecular orbitals (LUMO, LUMO+1, ...) as $\bra{\phi_{i_2}^{S(F)}}$. 
In both cases, $S(F)$ describes the total spin and the spin projection value of the final state $F$. 
The coefficients $\tilde{c}_{i_2,j_2}^{a_2}(F)$ and $\tilde{c}_{i_2}(F)$ determine the contribution of a CSF to the final state. We perform calculations employing RASSCF method \cite{MalmqvistJPhCm90} and use the same active orbitals for the treatment of the states $\ket{\Phi_\eSI}$ and $\bra{\Phi_F^{N_{\text{el}}-1}} $. With these approximations, the overlap integrals between neutral and singly-ionized eigenstates of a molecule are given by a linear combination of the overlap integrals between $N_{\text{el}}$- and $(N_{\text{el}}-1)$-electron CSFs. Such an overlap integral provide either a zero or a molecule orbital, which is occupied in a $N_{\text{el}}$-electron CSF and unoccupied in a $(N_{\text{el}}-1)$-electron CSF. Thus, a Dyson orbital can be represented as a linear combination of molecular orbitals.

\section{Imaging photo-induced dynamics in pentacene} \label{sec:dynmaicsPentacene}
Based on the previous considerations, we present an example for imaging excited-state dynamics in aligned isolated pentacene molecules. The molecular alignment can be achieved either for gas-phase molecules or molecular films adsorbed on substrates. We recently observed that photo-excited dynamics of pentacene molecules in a top pentacene layer of a bilayer pentacene films adsorbed on an Ag(110) surface behave similarly to the dynamics of an isolated pentacene \cite{baumgartner2022ultrafast}. These dynamics were probed using time-resolved photoelectron momentum microscopy on a time scale of several hundreds of femtoseconds. This experiment established conditions for time-resolved photoelectron microscopy at free-electron lasers, which makes it feasible to perform such experiments at much shorter timescales with brilliant and ultrashort light pulses from free-electron laser sources. 

In our study, we assume that coherent electron dynamics in pentacene were excited by a pump pulse via an optical excitation from the ground state. Since direct optical transitions from the ground state to triplet excited states are dipole forbidden, we truncate our considerations to spin-singlet excited states.

We calculate the vertical excited spin-singlet states of pentacene with MOLCAS \cite{https://doi.org/10.1002/jcc.24221_Molcas} using the RASSCF method \cite{MalmqvistJPhCm90}, with a CC-PVDZ basis set for the atoms \cite{dunning1989a_CC-PVDZ, pritchard2019a_BSE}. The calculation of the first excited states of pentacene is converged with an active space of 22 orbitals, containing 11$\pi$ and 11$\pi^*$ orbitals. Including more orbitals in the active space does not lead to significant changes in the energy spectrum and the structure of the eigenstates. The $\pi$ orbitals contain a maximum of one hole, while the $\pi^*$ orbitals have a maximum of one electron. The energies of the first four excited states and the changes in the occupation of the orbitals (compared to the HF ground state) are shown in Fig.~\ref{fig:pentaceneExcitedStates}. We find that the third and the fourth excited singlet states are dark states and the next bright singlet states are energetically highly separated from the first and the second excited state. Thus, one would be able to create a coherent superposition of just two excited states of a pentacene in a possible experiment. 
When the coherent superposition of the two excited states of pentacene is created, its electron density starts to oscillate in time. The oscillation period is determined by the energy difference between the excited states. 
The energy difference and the electronic wavefunctions of the bright excited states obtained in our calculation agree well with the experimental and theoretical results in previous studies \cite{heinecke1998laser,grimme2003substantial,Coto2015}. Also, the agreement with a more accurate CASPT2 calculation \cite{Coto2015} justifies our chosen theoretical level of single excitation in the configuration space. 
We obtain that the first excited state is predominantly characterized by the CSF obtained by the excitation of an electron from the HOMO to the LUMO. 
The second excited state is predominantly characterized by a linear combination of the CSF obtained by the excitation of an electron from the HOMO to the LUMO+2 and the CSF obtained by the excitation of an electron from the HOMO-2 to the LUMO.

\begin{figure}
    \centering
    \includegraphics[width=0.47\textwidth]{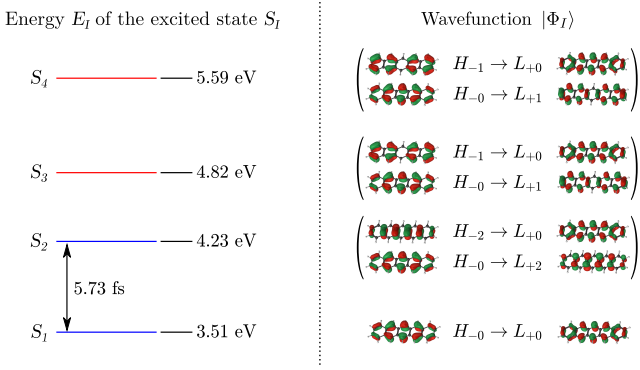}
    \caption{Energy level diagram of the first four vertical spin-singlet excited states of pentacene (left), with bright states in blue and dark states in red. A wave-packet in a coherent superposition of the first two excited states has an oscillation period of $5.73 \ \text{fs}$. The corresponding orbital excitations contribute to the wave function of the states (right), while the arrow means an excitation from the left (HOMO-i) to the right (LUMO+j) orbital. For the visualization of the orbitals, we used the software LUSCUS \cite{kovavcevic2015luscus}}
    \label{fig:pentaceneExcitedStates}
\end{figure}

The energy difference between the two excited states results in a oscillation period of the electron density of $T=5.73 \ \text{fs}$. 
To simplify our considerations, we can neglect nuclear motions that should affect electron dynamics on longer time scales. 
With \rEq{eq:cohState} the time evolution of the wave-packet after the excitation by the pump pulse at $t_0=0$ is
\begin{align} \label{eq:initialState}
    \ket{\Psi(t)} &= C_{1} e^{-i E_{1} t} \ket{\Phi_1} + C_{2} e^{-i E_{2} t} \ket{\Phi_2}, \\
    \ket{\Phi_1} &= \ket{\phi_{H}^{L,0(0)}}, \\
    \ket{\Phi_2} &= \frac{1}{\sqrt{2}} \left( \ket{\phi_{H}^{L+2,0(0)}} - \ket{\phi_{H-2}^{L,0(0)}} \right),
\end{align}
with $|C_1|^2+|C_2|^2=1$. The coefficients $C_1$ and $C_2$ depend on a particular process, which brought a molecule into a coherent superposition of electronic states. Our results do not depend on the particular values of the coefficients $C_1$ and $C_2$ and we assume an equal population of both excited states, $C_1 = C_2 = \frac{1}{\sqrt{2}}$. 
The orbital index $H-i$ denotes the $i$-th orbital below the HOMO and $L+j$ is the $j$-th orbital above the LUMO. The photo-excitation leads to a time-dependent change in the electron density relative to the ground state density
\begin{equation}
    \rho_{ex} (\br, t) = \rho_1(\br, t) - \rho_0(\br),
\end{equation}
where $\rho_0(\br) = \bra{\Phi_0} \hat{\psi}^\dagger(\br) \hat{\psi}(\br) \ket{\Phi_0}$ is the electron density of the ground state of pentacene described by the wave function $\ket{\Phi_0}$. The electron density for the pentacene after the photo-excitation by a pump pulse is $\rho_1(\br, t) = \bra{\Psi(t)} \hat{\psi}^\dagger(\br) \hat{\psi}(\br) \ket{\Psi(t)}$. 
We evaluate the electron density change and obtain
\begin{align} \label{eq:specificExDens}
    \begin{split}
        \rho_{ex} (\br, t) = &
        \left| 
            C_1^* e^{i E_1 t } \phi_{L}(\br) 
            + \frac{C_2^*}{\sqrt{2}} e^{i E_2 t} \phi_{L+2}(\br)
        \right| ^2
        \\ 
        & + \frac{|C_2|^2}{2} |\phi_{L}(\br)|^2 - \frac{|C_2|^2}{2} |\phi_{H}(\br)|^2\\
        & -
        \left| 
            C_1^* e^{i E_1 t } \phi_{H}(\br) 
            - \frac{C_2^*}{\sqrt{2}} e^{i E_2 t } \phi_{H-2}(\br)
        \right| ^2
        ,
    \end{split}
\end{align} 
where $\phi_{H-i}(\br)$ and $\phi_{L+j}(\br)$ are the wavefunctions of the i-th orbital below the HOMO and the j-th orbital above the LUMO. 
The electron density change is shown in Fig.~\ref{fig:excitonDensity} at different times during the oscillation period. 
The negative (blue) part is the reduction of the electron density compared to the ground state density and the positive (yellow) part is the increase of the electron density compared to the ground state density. 
The time-dependent part of the electron density oscillates in time as $\cos[(E_1 - E_2)t]$. 
From $t=0$ to $t=\frac{T}{2}$, the photo-induced negative charge flows from the top left and bottom right of the molecule to the top right and bottom left. The photo-induced positive charge flows in the opposite direction. The spatial distribution of the electron density change at $t=\frac{T}{2}$ can be mapped onto the electron density change at $t=0$ by a reflection in the yz-plane. 
At two times during the oscillation period, $t=\frac{T}{4}$ and $t=\frac{3T}{4}$, the photo-excited change in the electron density coincides and has reflection symmetry. 

\begin{figure}
    \centering
    \includegraphics[width=0.4\textwidth]{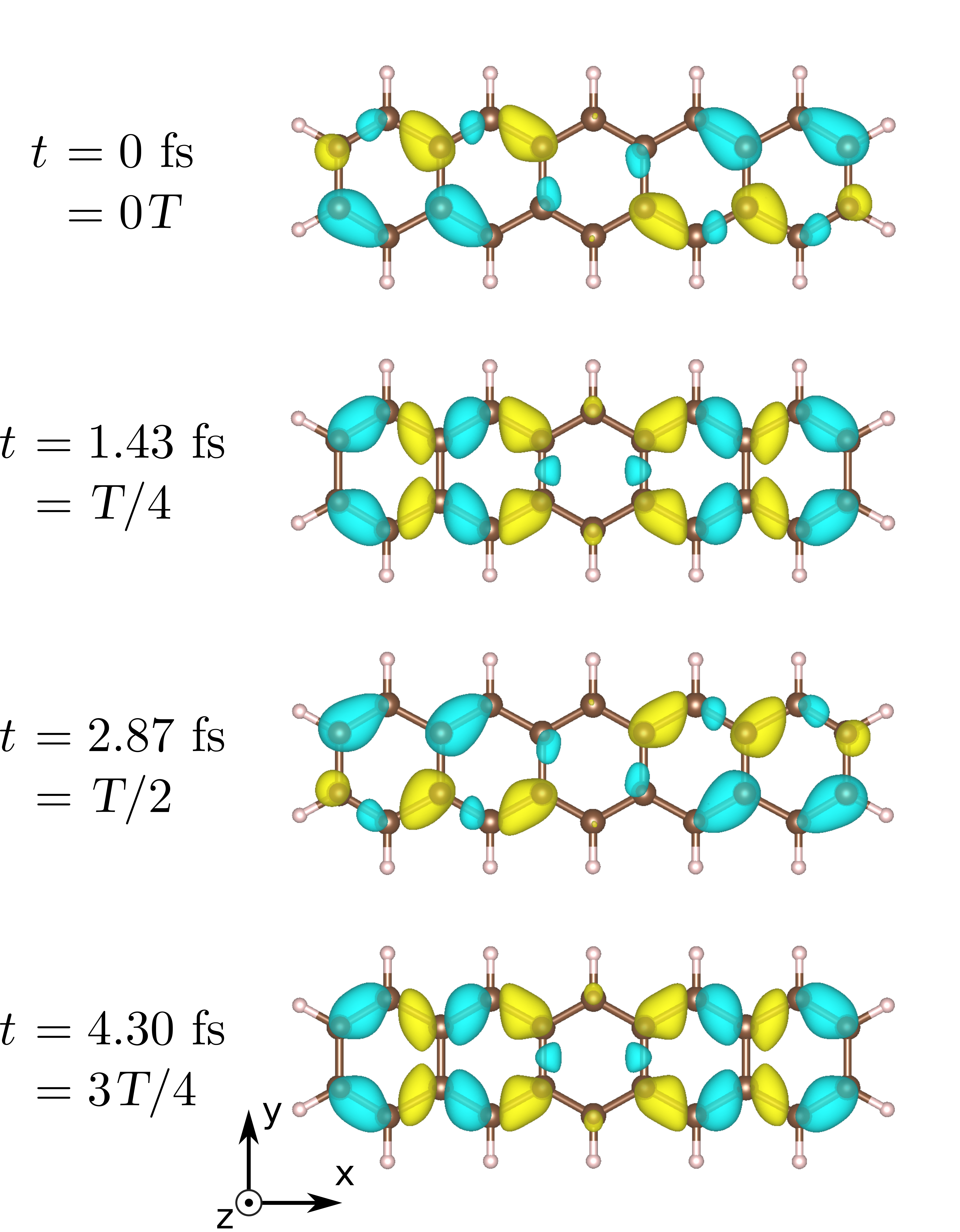}
    \caption{The time evolution of the electron-hole density after the excitation of the pentacene molecule to the first two excited states with the pump pulse at $t_0=0 \text{ fs}$. The blue isosurface shows the hole density (negative value) and the yellow the electron density (positive value). For the visualization we used the software VESTA \cite{MommaJAC11}.}
    \label{fig:excitonDensity}
\end{figure}
\renewcommand{\arraystretch}{2.0}
\begin{table*}
    \centering
    \begin{ruledtabular}
    \begin{tabular}{cccc}
        F & $E_F$ & $\Omega_{F}$ & $\bra{\Phi_F^{N_{\text{el}}-1}} $\\ \hline
        1 & $5.0 \ \text{eV}$ & $98.9 \ \text{eV}$  & $-0.95 \bra{\Phi_{H}^{1/2(1/2)}}$ \\
        2 & $6.7 \ \text{eV}$ & $97.2 \ \text{eV}$  & $-0.94 \bra{\Phi_{H-2}^{1/2(1/2)}}$ \\
        3 & $7.5 \ \text{eV}$ & $96.4 \ \text{eV}$  & $-0.83 \bra{\Phi_{H,H}^{L,1/2(1/2)}} + 0.31\bra{\Phi_{H-4}^{1/2(1/2)}}$ \\
        4 & $8.7 \ \text{eV}$ & $95.2 \ \text{eV}$  & $-0.62\bra{\Phi_{H,H}^{L+1,1/2(1/2)}} + 0.5 \bra{\Phi_{H-1,H}^{L,1/2(1/2)}}_{udu}$
        $+ 0.28 \bra{\Phi_{H-1,H}^{L,1/2(1/2)}}_{uud}$\\
        5 & $9.6 \ \text{eV}$ & $94.3 \ \text{eV}$  & $0.70 \bra{\Phi_{H-1,H}^{L,1/2(1/2)}}_{uud} - 0.45 \bra{\Phi_{H-1,H}^{L,1/2(1/2)}}_{udu}$ \\
        6 & $9.7 \ \text{eV}$ & $94.2 \ \text{eV}$    & $-0.59 \bra{\Phi_{H,H}^{L+2,1/2(1/2)}} + 0.62 \bra{\Phi_{H-2,H}^{L,1/2(1/2)}}_{udu} $ 
        $- 0.28 \bra{\Phi_{H-2,H}^{L,1/2(1/2)}}_{uud}$ \\
    \end{tabular}
    \end{ruledtabular}
    \caption{\label{tab:fs} First six final states $\bra{\Phi_F}$ of ionized and excited pentacene, which result in a nonzero Dyson orbital $\phi_F^D(\br) = \bra{\Phi_F^{N_{\text{el}}-1}} \hat{\psi}(\br) \ket{\Psi(t)}$. The corresponding energy is $E_F$ and $\Omega_{F} = E_F - \left<E \right> - \omega_{\text{in}} + \epsilon_e$ is the central energy in the photoelectron probability in \rEq{eq:photoElProp} for the given final state. The right column shows the contributing configuration state functions of the final state. The indices $udu$ and $uud$ distinguish two different CSFs that result in the total spin of 1/2 and spin projection of +1/2.}
\end{table*}

\begin{figure}
    \centering
    \includegraphics[width=0.42\textwidth]{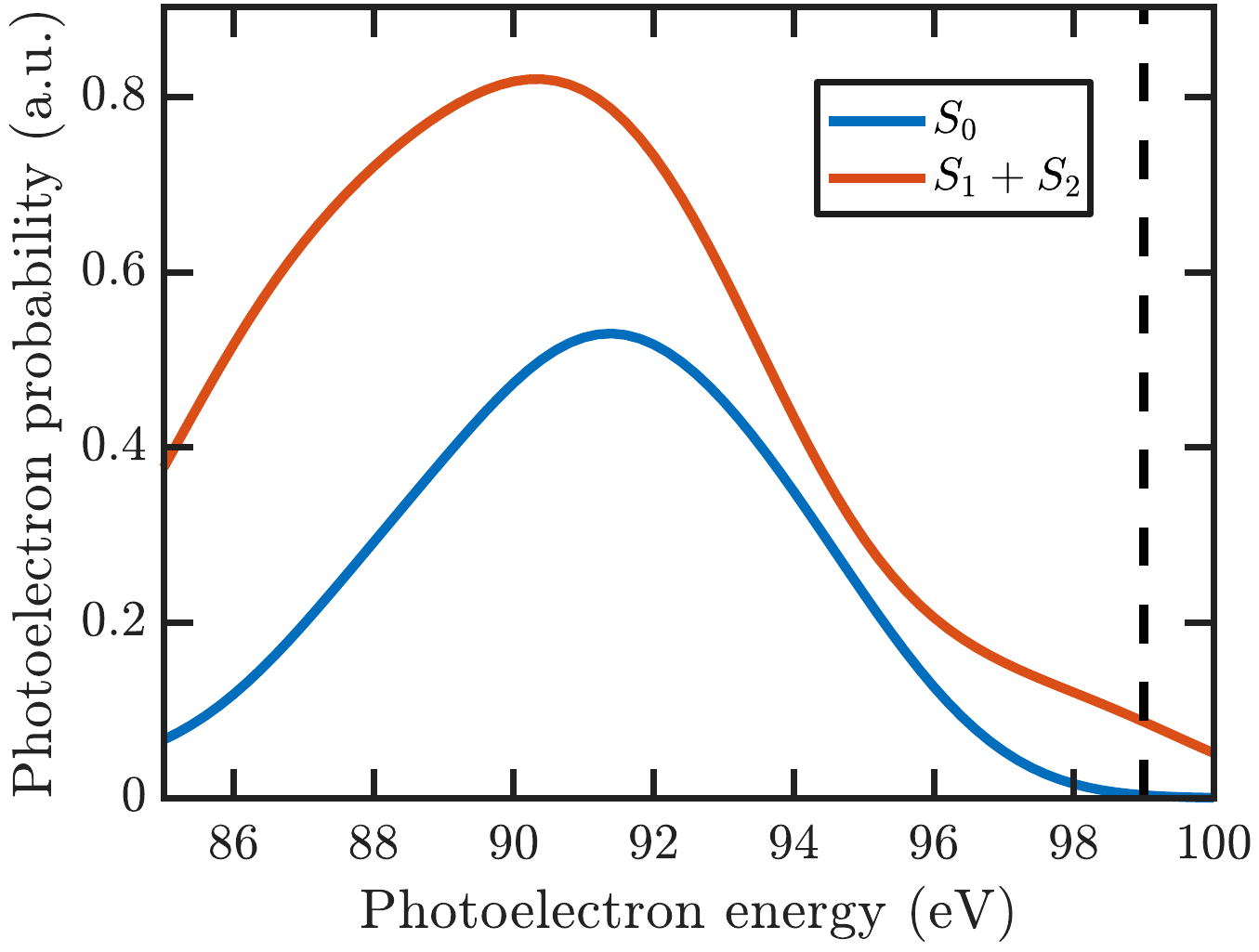}
    \caption{Angle-integrated photoelectron spectra for photoelectrons detached either from the ground state $S_0$ or in the coherent superposition of the excited state $S_1$ and $S_2$. The black dotted line shows the contribution at the photoelectron energy $\epsilon_e = 99 \ \text{eV}$. The probe-pulse duration is $0.5 \ \text{fs}$ and has a central photon energy of $100 \ \text{eV}$.}
    \label{fig:angleIntSpectra}
\end{figure}

We assume that an XUV probe pulse with the central energy of $100 \ \text{eV}$ and $500 \ \text{as}$ duration photoionizes the molecule at the pump-probe time delay $t_p$. The probe-pulse is linearly polarized along the $y$-direction, which is perpendicular to the nodal plane of the molecule. For the calculation of the final states after the photoionization by the probe pulse, we apply the same method as in the calculation of the excited states of pentacene, but with an additional hole in the 11 active $\pi$ orbitals. 
Tab.~\ref{tab:fs} shows the first six final states of pentacene with the lowest energy that can be reached by ionizing the initial state in \rEq{eq:initialState}. 

If the photoionization was triggered by a pulse with a narrow bandwidth, the photoelectron probability as a function of energy would consist of clear spectral lines centred at energies $\Omega_{F,I}=E_F - E_{I} - \omega_{\text{in}} + \epsilon_e$. Thus, the photoelectron momentum distributions would be clearly assigned to a certain Dyson orbital. In the case of the ultrashort probe pulse, each spectral line has the spectral band width of $5.16 \ \text{eV}$ and an individual Dyson orbital contributing to the momentum distribution at a given energy is no longer distinguished.

Let us first consider the angle-integrated photoelectron spectra. In a real experiment, a probe pulse would interact with a number of pentacene molecules, some of which would be excited by a pump pulse, and some of which would remain unexcited. Thus, we first estimate, how the photoionization from unexcited pentacene molecules would affect the photoelectron spectra from pentacene molecules in the excited state. In Fig.~\ref{fig:angleIntSpectra}, we show photoelectron spectra from the pentacene molecules being in the ground state $S_0$ and from pentacene molecules being in the coherent superposition of the excited state $S_1$ and $S_2$ ($\ket{\Psi(t)}$ in \rEq{eq:initialState}). We obtain that the contribution of photoelectrons detached from unexcited molecules is negligible at energies higher than 97 eV. Thus, photoelectron spectra at energies higher than 97 eV are beneficial to study the dynamics of photoexcited pentacene molecules, since they would not contain a background signal from the unexcited molecules. 

We find a remarkable result that the angle-averaged photoelectron spectra from photo-excited pentacene do not vary with the probe-pulse arrival time although the photo-induced charge distribution experiences considerable oscillations in time (see Fig.~\ref{fig:excitonDensity}). This is because the time-dependent part of the electronic density change is highly symmetric in space. As a consequence, the photoelectron momentum distributions are also highly symmetric and the temporal dependence of the signal averages out after the angle integration. The angle-integrated spectra do not reveal information about the time-dependent photo-induced charge oscillations.
 
We now demonstrate that photoelectron momentum microscopy is indeed quite sensitive to photo-induced electron dynamics. 
We consider photoelectron momentum maps (PMMs) of a photoelectron with a fixed kinetic energy of $\epsilon_e = 99 \ \text{eV}$. 
PMMs are constructed by a two-dimensional projection of the three-dimensional photoelectron momentum distribution for a constant energy $\epsilon_e$ (hemispherical cuts). 
Based on \rEq{eq:photoElProp}, the photoelectron probability resulting from a particular final state is only time-dependent, if the final state has a nonzero overlap with both eigenstates in the initial state, $\bra{\Phi_F} \hat{\psi}(\br) \ket{\Phi_1} \neq 0$ and $\bra{\Phi_F} \hat{\psi}(\br) \ket{\Phi_2} \neq 0$. At the chosen kinetic energy, three final states with the lowest energies, with the indices 1, 2 and 3 in Table~\ref{tab:fs} contribute to the PMM. Thereby, among these three final states, only the final state 1 contributes to a time-dependent signal and the other two final states 2 and 3 contribute to a time-independent background. The first final state results in a broad spectral line centered approximately at the energy $\epsilon_e = 99 \ \text{eV}$.

We derive the explicit expressions for the Dyson orbital that determine the angle-resolved photoelectron spectra at $\epsilon_e = 99 \ \text{eV}$. For the first three final states, we obtain
\begin{align} 
\begin{split}
    \phi_{F_1}^D (\br) &= \bra{\Phi_{F_1}} \hat{\psi}(\br) \ket{\Psi(t)} \\
    &= 
    -C_1 e^{-i E_1 t_p } \frac{0.95}{\sqrt{2}} \phi_L (\br)  \chi_{\da}\\
    & \hphantom{=}-C_2 e^{-i E_2 t_p } \frac{0.95}{2} \phi_{L+2} (\br) \chi_{\da}, 
    \label{eq:dys1}
\end{split}\\
\begin{split}
    \phi_{F_2}^D (\br) &= \bra{\Phi_{F_2}} \hat{\psi}(\br) \ket{\Psi(t)} \\
    &= C_2 e^{-i E_2 t_p } \frac{0.94}{2} \phi_L (\br) \chi_{\da}, \label{eq:dys2} 
\end{split}\\
\begin{split}
    \phi_{F_3}^D (\br) &= \bra{\Phi_{F_3}} \hat{\psi}(\br) \ket{\Psi(t)}  \\
    &= - C_1 e^{-i E_1 t_p } \frac{0.83}{\sqrt{2}} \phi_H (\br) \chi_{\da}. \label{eq:dys3}
\end{split}
\end{align}

\begin{figure*}
    \centering
    \includegraphics[width=0.9\textwidth]{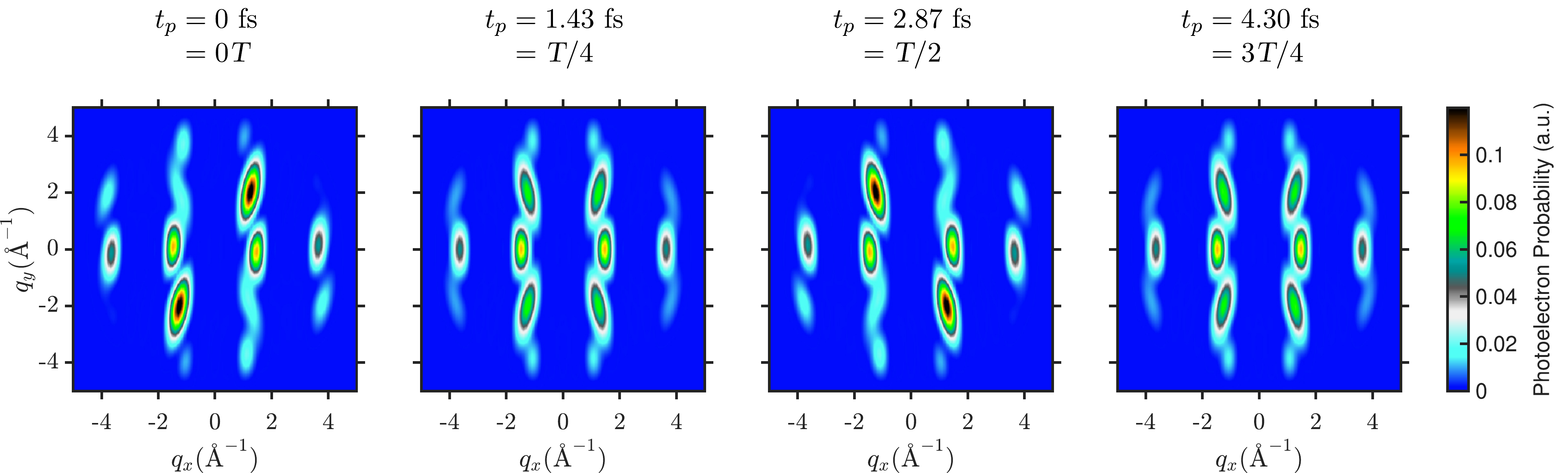}
    \caption{Time-dependent photoelectron momentum maps for photoelectrons emitted from the excited pentacene at different probe-pulse arrival times $t_p$. The probe-pulse duration is $0.5 \ \text{fs}$ and the central photon energy is $100 \ \text{eV}$. The momentum maps are shown at the photoelectron energy of $\epsilon_e = 99 \ \text{eV}$.}
    \label{fig:trPmms}
\end{figure*}
These three Dyson orbitals in Eqs.~(\ref{eq:dys1}) - (\ref{eq:dys3}) enter the photoelectron probability in \rEq{eq:photoElProp} at the photoelectron energy $\epsilon = 99 \ \text{eV}$
\begin{widetext}
\begin{align}
\begin{split} \label{eq:photoeElPro99}
    P (\bq, t_p) \underset{\epsilon_e = 99 \ \text{eV}}{\propto} 
    &\left| \bep_{\text{in}} \cdot \bq \right|^2 \left[ e^{-(\Omega_{F_1} - \epsilon_e )^2\frac{\tau_p^2}{4\ln2}} 
    \left| C_1 e^{-i E_1 t_p } \frac{0.95}{\sqrt{2}} \mathcal{F} (\phi_L (\br))(\bq) \right. \right.
    \left. \left. +  C_2 e^{-i E_2 t_p } \frac{0.95}{2} \mathcal{F} (\phi_{L+2} (\br))(\bq) \right|^2 \right] \\
    +  &\left| \bep_{\text{in}} \cdot \bq \right|^2 \left[ e^{-(\Omega_{F_2} - \epsilon_e )^2\frac{\tau_p^2}{4\ln2}}
    \left| C_2 \frac{0.94}{2} \mathcal{F} (\phi_L (\br))(\bq)\right|^2 \right.
    + \left. e^{-(\Omega_{F_3} - \epsilon_e )^2\frac{\tau_p^2}{4\ln2}} \left| C_1 \frac{0.83}{\sqrt{2}} \mathcal{F} (\phi_{H} (\br))(\bq) \right|^2 \right],
\end{split}
\end{align}
\end{widetext}
where $\mathcal{F} (\phi (\br))(\bq) = (2 \pi)^ {-3/2} \int \text{d}r^3 e^{i \bq \cdot\br} \phi (\br)$ denotes the Fourier transform of an orbital $\phi (\br)$. 
The resulting time-resolved PMMs shown in Fig.~\ref{fig:trPmms} vary strongly with time. 
The PMMs at $t_p=0$ and $t_p=\frac{T}{2}$ are the point reflection image of each other and have the same symmetry as the changes of the electron density at these times (see Fig.~\ref{fig:excitonDensity}). 
The PMMs at $t_p=\frac{T}{4}$ and $t_p=\frac{3T}{4}$, when the electron density coincides, are equal and have reflection symmetry. 
The most pronounced oscillating peaks are located in the PMMs at the momenta $q_x = \pm 1.26\ \angstrom^{-1}$ and $q_y = \pm 1.97\ \angstrom^{-1}$. 
According to \rEq{eq:photoeElPro99} the time-dependent part of the PMMs arises from the Fourier transform of LUMO and LUMO$+2$ orbitals. It is directly related to the negatively-charged part of the induced electron-density change (yellow in Fig.~\ref{fig:excitonDensity}), which is also specified by the LUMO and LUMO$+2$ orbitals (see \rEq{eq:specificExDens}). Thus, the time-dependent features in the PMMs encode the information about the symmetry of the photo-induced electron density. Due to a high contrast of a signal and a low background, these features can be detected in an experiment even if the signal is weak. Due to the high spatial resolution, PMMs resolve oscillations of the electron density within the molecule.

\begin{figure*}
    \centering
    \includegraphics[width=0.95\textwidth]{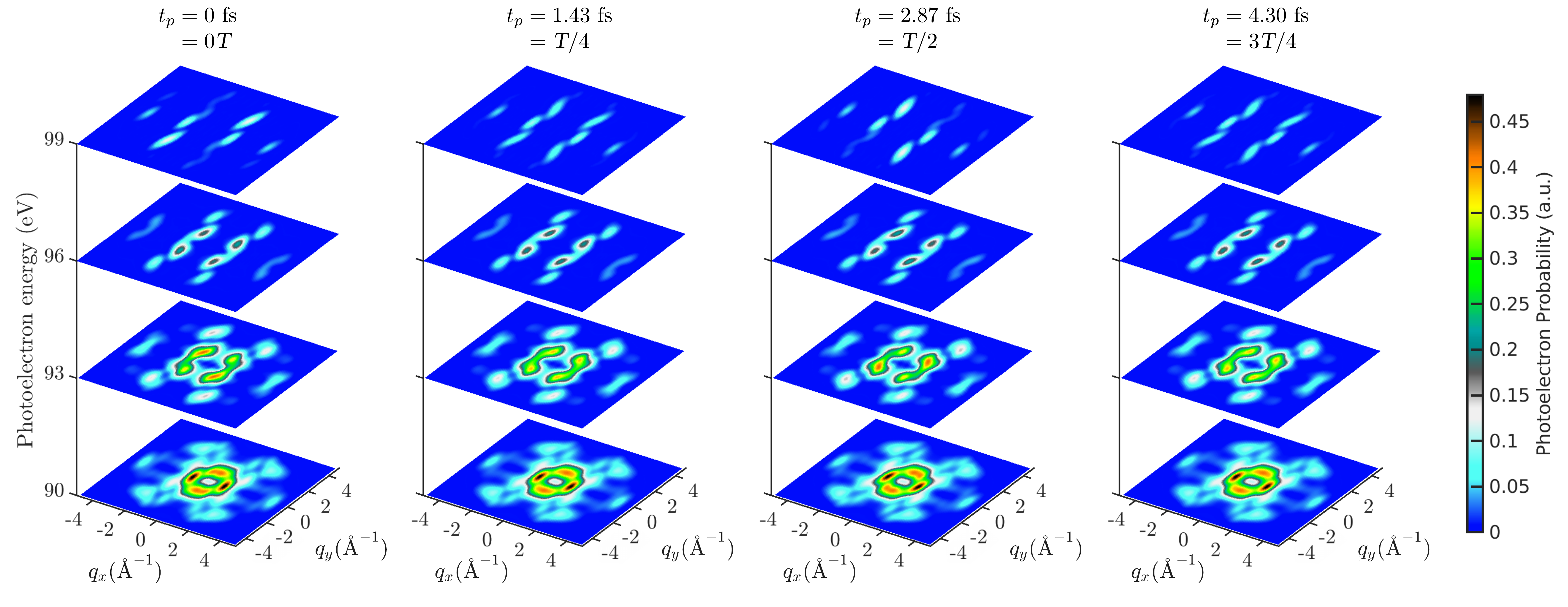}
    \caption{Time-dependent photoelectron momentum maps for photoelectrons emitted from the excited pentacene for different photoelectron energies $\epsilon_e$. Each set of momentum maps has a different probe-pulse arrival time $t_p$. The probe-pulse duration is $0.5 \ \text{fs}$ and the central photon energy is $100 \ \text{eV}$.}
    \label{fig:angle3dSpectra}
\end{figure*}

In Ref.~\cite{Popova-GorelovaPRA16, app8030318}, we demonstrated that that a general time- and momentum-resolved signal from a coherently evolving electronic wave packet can be sensitive to electron currents within a sample. This means that a momentum-resolved signal can be even different, when the electron density coincides, but the charge flow direction differs. This applies to the electronic state in Fig.~\ref{fig:excitonDensity} at times $t_p=\frac{T}{4}$ and $t_p=\frac{3T}{4}$. However, we do not observe this effect in the considered case, since it also averages out due to the high symmetry of the photo-induced charge distribution.

Finally, we investigate PMMs and their temporal dependence at different photoelectron energies in Fig.~\ref{fig:angle3dSpectra}. 
As follows from the angle-integrated spectra in Fig.~\ref{fig:angleIntSpectra}, the overall intensity of the PMMs increases at lower photoelectron energies. 
The PMMs at energy $\epsilon_e = 90 \ \text{eV}$ and at energy $\epsilon_e = 96 \ \text{eV}$ remain almost constant in time, whereas the PMMs at $\epsilon_e = 93 \ \text{eV}$ and at $\epsilon_e = 99 \ \text{eV}$ show pronounced temporal dependence. Although the absolute change of the signal in time is approximately the same in the two former cases, the overall contrast is much better for the PMMs at $\epsilon_e = 99 \ \text{eV}$ due to a lower time-constant background. This temporal behavior of the signal can be explained by the character of photo-induced changes in the electron density. These changes are due to the electrons photo-excited into orbitals that were unoccupied in the ground state and holes in the orbitals that were doubly occupied in the ground state. If an electron, which evolves in the photo-excited orbitals, is emitted by a probe pulse, then it directly carries the information about the negatively-charged oscillations of the electron density. If an electron is indeed emitted from lower lying orbitals, where an electron hole evolves, then it is an "indirect probe" of the evolving hole. 
Thus, the signal at the photoelectron energy of 99 eV results from the emission of electrons evolving in the photo-excited orbitals and is the most advantageous to obtain a time-resolved signal with a high contrast.

\begin{figure*}
    \centering
    \includegraphics[width=\textwidth]{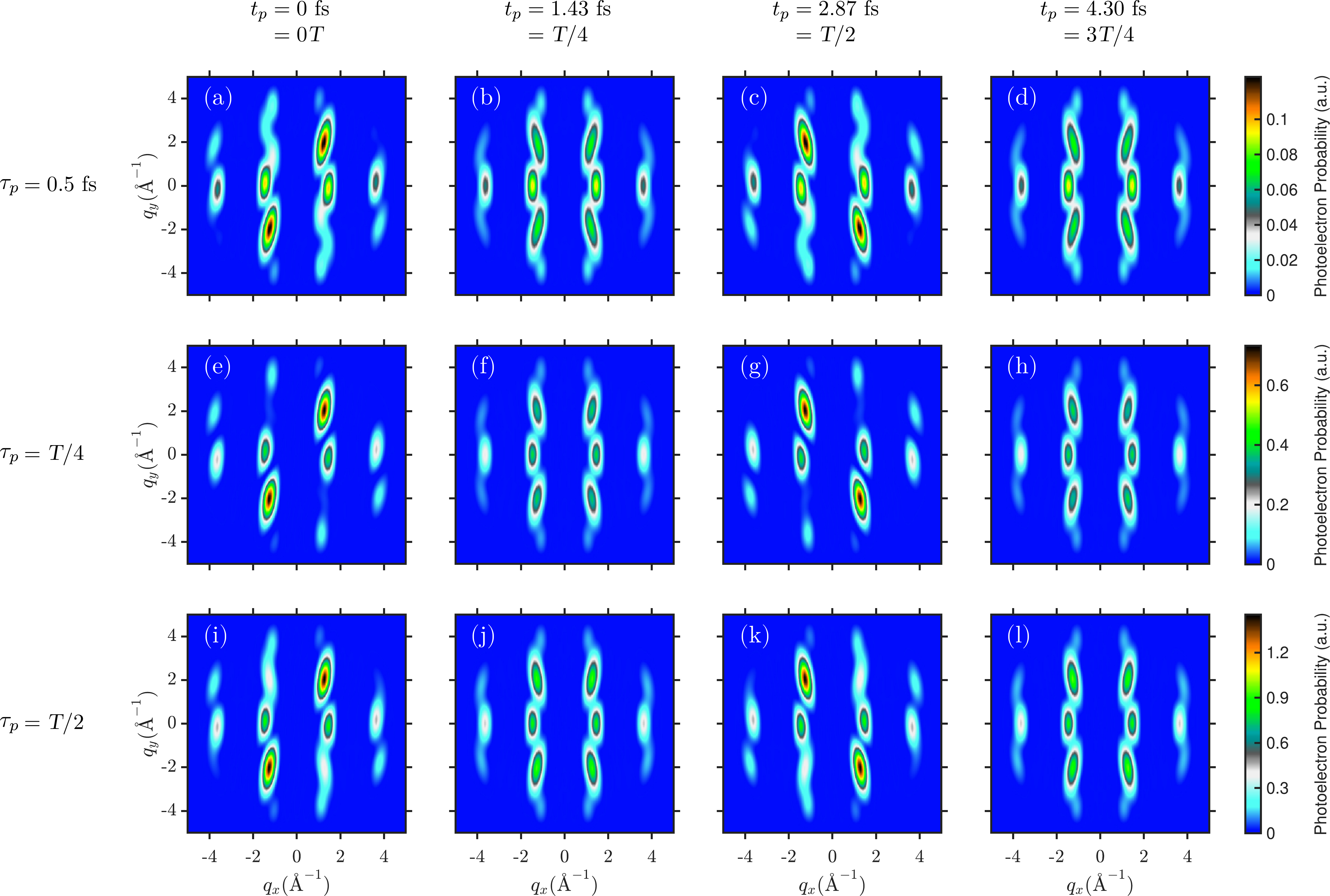}
    \caption{Time-dependent photoelectron momentum maps (PMMs) for photoelectrons emitted from the excited pentacene. In each column the PMMs are shown for different probe-pulse arrival times (a),(e),(i) $t_p= 0T$, (b),(f),(j) $t_p= T/4$, (c),(g),(k) $t_p= T/2$, (d),(h),(l) $t_p= 3T/4$. In the different rows the PMMs are shown for different probe-pulse durations (a)-(d) $\tau_p=0.5 \ \text{fs}$, (e)-(h) $\tau_p=T/4$, (i)-(l) $\tau_p=T/2$. The probe pulse has a central photon energy of $100 \ \text{eV}$. The momentum maps are obtained by averaging the signal over the range of $\epsilon_e = 98.5 \ \text{eV}$ to $\epsilon_e = 99.5 \ \text{eV}$ to simulate the energy resolution of 1 eV.}
    \label{fig:trPmmsInt}
\end{figure*}

In a real experiment, PMMs would be obtained with limited energy and time resolution. We simulate the PMMs in the case of limited energy resolution by averaging the photoelectron probability over the corresponding energy interval. We show the PMMs at the photoelectron energy of 99 eV assuming the energy resolution of 1 eV in Figs.~\ref{fig:trPmmsInt}(a)-(d), which are obtained by averaging the signal over the range of 98.5 eV to 99.5 eV. We obtain that the momentum distributions are almost unaffected by the limited energy resolution.

The temporal resolution of the experiment is determined by the duration of the probe pulse and the timing jitter between the probe and the pump pulses. If the probe-pulse duration cannot be assumed to be much shorter than the oscillation period, Eq.~(\ref{Eq_long_pulse_duration}) must be used for the calculation instead of Eq.~(\ref{eq:photoElProp}). We calculate the signal with Eq.~(\ref{Eq_long_pulse_duration}) at different probe-pulse durations $\tau_p$. We also assume the limited energy resolution of 1 eV. As shown in Figs.~\ref{fig:trPmmsInt}(e)-(h) and Figs.~\ref{fig:trPmmsInt}(i)-(l), the temporal contrast of the resulting signal is quite high even, if the probe-pulse duration is as large as a quarter or half of the oscillation period. 

In our calculation, we neglected electron-nuclear coupling in the description of charge migration since the primary goal of our study is to investigate how pure electron dynamics on their natural time scale can be measured. Let us discuss, which consequences the coupling of electron dynamics to nuclear dynamics would have for the signal. The coupling to nuclear motions would indeed lead to decoherence of electron dynamics and should cause the decrease of the oscillation amplitude \cite{PhysRevA.95.033425, doi:10.1063/1.4996505, PhysRevLett.121.203002, Matselyukh2022}. Former studies of coupled electron-nuclear dynamics of similar molecules such as norbornadiene cations \cite{doi:10.1063/1.4965436} or benzene \cite{doi:10.1021/jz502493j} have demonstrated that coherent electron motion can last up to 20 fs in these molecules. It is expected that coherent electronic motion in polycyclic aromatic hydrocarbons, which also include pentacene, would experience similar time scales \cite{doi:10.1021/jz502493j}. Since 20 fs corresponds to about five cycles of oscillation in our case, our assumption that charge migration is coherent during at least one full oscillation period is robust. At longer time scales, the decay of the amplitude of the periodic charge oscillations due to decoherence would result in the decrease of the time-dependent changes in the PMMs. This would lead to a loss of temporal contrast of the signal, which means that the signal at different times would become less distinguishable from the time-independent part of the signal (see Fig.~\ref{fig:angle3dSpectra}). This could serve even as an advantage, since such an experiment would reveal the transition of coherent dynamics to decoherent dynamics.

\section{Conclusions}

We showed theoretically how time-resolved photoelectron momentum microscopy can be employed to image photo-induced coherently evolving electron dynamics in a pentacene molecule. 
The assumed XUV probe pulse of 500 as duration provides the temporal resolution, which is sensitive to pure intramolecular electronic dynamics. 
The momentum-resolved signal revealed details of electron dynamics with \r Angstrom spatial resolution. 
Although the angle-averaged photoelectron spectra were not sensitive to electron dynamics, momentum-resolved spectra provided a highly contrast signal evolving in time. 
At high photoelectron energies, the signal is provided by the electrons emitted from photo-excited orbitals and is not affected from a possible contribution of photoelectrons emitted by molecules that were not photo-excited. 
Such photoelectron momentum maps are directly related to the negatively-charged part of the photo-induced changes in the electron density and is sensitive to its symmetry in real space. 
These findings demonstrate that attosecond photoelectron momentum microscopy has a big potential to become a novel technique to measure charge migration and electron coherence processes in exciton transport in organic semiconductors with unprecedented time and spatial details.

\section{Acknowledgements}
We gratefully acknowledge the funding of the Volkswagen Foundation.

\end{document}